\documentstyle[axodraw,preprint,aps]{revtex}

\begin{document}
\tighten
\draft


\title{Proton Decay in Minimal Supersymmetric SU(5)}

\author{Borut Bajc$^{(1)}$, Pavel Fileviez Perez
$^{(2)}$, and Goran Senjanovi\'c$^{(3)}$}

\address{~}

\address{$^{(1)}$ {\it J. Stefan Institute, 1001 Ljubljana, Slovenia}}
\address{$^{(2)}${\it Max-Planck Institut f\" ur Physik
(Werner Heisenberg Institut), F\" ohringer Ring 6,
80805 M\" unchen, Germany}}
\address{$^{(3)}${\it International Centre for Theoretical Physics,
Trieste, Italy}}

\maketitle

\begin{abstract}
We systematically study proton decay in the minimal supersymmetric
SU(5) grand unified theory. We find that although the available 
parameter space of soft masses and mixings is quite constrained, 
the theory is still in accord with experiment. 
\end{abstract}

\section{Introduction}

It has been known for more than ten years that the low energy
supersymmetry is tailor fit for grand unification: with the desert
assumption the gauge couplings of the supersymmetric standard model
unify at the single scale $M_{GUT}\approx 10^{16}$ GeV. Actually,
this was foreseen some twenty years ago
\cite{Dimopoulos:1981yj,Ibanez:yh,Einhorn:1981sx,Marciano:1981un}.
However, it was noticed almost immediately
that supersymmetric GUTs \cite{Sakai:1981pk,Weinberg:1981wj}
carry a potential catastrophe of new $d=5$ contributions to the
proton decay. This has been studied on and off for the last twenty
years (see for example \cite{Hisano:1992jj},\cite{Goto:1998qg})
with the culminating conclusion
\cite{Murayama:2001ur} that the minimal supersymmetric SU(5)
theory is actually ruled out precisely due to the $d=5$ proton decay.
To us ruling out the minimal theory is almost a death blow to the
idea of grand unification. It is hard enough to verify the predictions
of the minimal GUT; the extended versions of the theory unfortunately
stop being predictive. For example, the beauty of matter unification
and the naturalness of the see-saw mechanism \cite{Mohapatra:1981yp}
make a minimal SUSY SO(10) theory \cite{Aulakh:2000sn} more appealing.
However, this is a typical example of what we are saying: the theory
connects different mass scales, but does not predict them.

In view of the above it is of extreme importance to be completely
sure that the minimal SUSY GUT is ruled out. This has prompted us
to re-investigate this issue in gory detail. According to us, any
rumor of the death of the theory is somewhat premature. More
precisely, we study proton decay with arbitrary soft masses
and fermion and sfermion mixings and find out the following:
the model parameter space is quite constrained but not yet in
contradiction with experiment. In other words, the improved
measurements of proton decay will provide information about
the nature of supersymmetry breaking (i.e. the soft masses)
and the fermionic mass textures. This is the sector of the
theory completely orthogonal to grand unification and therefore
we advocate the point of view that proton decay is not yet a
good test of the generic properties of grand unification (here
we mean obviously the dimension 5 aspect of it). We should
stress here that the so called decoupling regime seems to be both 
necessary and sufficient to save the theory from being ruled out.

In short, although we follow \cite{Murayama:2001ur} in
accepting the decoupling of the first two generations of
sfermions, we cannot agree on this not being enough.
The point is that we know nothing about individual fermion
and sfermion mixings. Thus, proton decay simply limits these
parameters and, admittedly, the restrictions are quite severe.
In all honesty, it is hard to imagine a simple scenario of
SUSY breaking which could be in accord with our constraints.
However, a phenomenological study must always be separated
from theoretical bias and, phenomenologically speaking, the
theory is still alive.

\section{The minimal supersymmetric SU(5)}

Before starting any discussion of proton decay, one must enter
the subtle issue of defining a minimal SU(5) theory. Obviously,
a reasonable definition should be based on choosing a minimal
Higgs sector which contains an adjoint {\bf 24} and a pair of
{\bf 5} and ${\bf\bar 5}$ representations.

We will show at the end of the day 
that even this theory (as incomplete as it is) is not in
conflict with the proton decay experiment. In order to be as
general as possible we perform our calculations for
arbitrary values of the parameters of the theory.

In minimal SU(5) we can most generally write (in the renormalizable
limit) for the relevant terms in the superpotential of the Higgs and
Yukawa sectors

\begin{equation}
\label{wh}
W_H={m_\Sigma\over 2}Tr\Sigma^2+{\lambda\over 3}Tr\Sigma^3
+\eta\bar 5_H\Sigma 5_H + m_H \bar 5_H 5_H\;,
\end{equation}

\begin{eqnarray}
\label{wy}
W_Y=&&5_H 10^T Y^U 10 + \bar 5_H 10^T Y^D \bar 5 \;,
\end{eqnarray}

\noindent
where $\Sigma$ is the SU(5) adjoint, $5_H$ and $\bar 5_H$ are the
Higgs fundamental and anti-fundamental superfield representations,
the 10 and $\bar 5$ refer to the three generations of matter 
superfields, and $Y's$ are $3\times 3$ Yukawa matrices.

In the supersymmetric standard model language the Yukawa sector can be
rewritten as

\begin{eqnarray}
\label{wmssm}
W_Y=&&HQ^TY_Uu^c+\bar H Q^TY_Dd^c+\bar H {e^c}^TY_EL\nonumber\\
&+&TQ^T\underline AQ+T{u^c}^T\underline B e^c+\bar TQ^T\underline CL+
\bar T {u^c}^T\underline Dd^c\;,
\end{eqnarray}

\noindent
where except for the heavy triplets $T$ and $\bar T$ the rest are 
the MSSM superfields in the usual notation. The generation matrices 
$Y_{U,D,E}$ and $\underline A$, $\underline B$, $\underline C$ and
$\underline D$ can in principle be arbitrary. In the minimal SU(5) 
defined above one finds the 
usual relations $\underline A=\underline B=Y_U=Y_U^T$, and
$\underline C=\underline D=Y_D=Y_E$ at the GUT scale.
The above definition of minimality implies no new structure at all
energies up to $M_{Pl}$. On the other hand, the lepton-down quark
relations can be easily corrected by higher dimensional operators
without introducing any new field at $M_{GUT}$. We postpone the
discussion of higher dimensional operators for the summary and
outlook.

As we mentioned before, we do not assume any specific values for
the soft mass matrices of squarks and sleptons. However, as
emphasized clearly in \cite{Murayama:2001ur}, we can not have
all three generations of squarks contribute to the proton decay.
The simplest direction to take as \cite{Murayama:2001ur} already
did, is to assume the so called decoupling limit for the sfermions:
the first two generations have a mass of order $10$ TeV, thus
effectively decoupling from the rest, while the third is of order $1$
TeV \cite{Pomarol:1995xc,Dvali:1996rj,Cohen:1996vb}. 
This is still in accord with 
naturalness constraints and the limits from flavour violation in
neutral current phenomena suggest small mixings with the first 
two generations of fermions. We will see later that it is possible 
to make the proton decay be in agreement with experiment, again for 
some combinations of such mixings being small. 

With this in mind we allow the mass diagonalization matrices to be
different for particles and sparticles. For the fermions we
have

\begin{eqnarray}
U^T Y_U U_c&=&Y_U^d\;,\nonumber\\
D^T Y_D D_c&=&Y_D^d\;,\\
E_c^T Y_E E&=&Y_E^d\;,\nonumber
\end{eqnarray}

\noindent
where $X$ ($X_c$) is the unitary matrix that rotates the
fermion $x$ ($x^c$) from the flavour to the mass basis. The only
combination we know from low-energy experiments is
$U^\dagger D=V_{CKM}$ (and a similar one in the lepton sector,
$N^\dagger E=V_l$, the leptonic mixing matrix).

Similarly, the unitary matrices $\tilde X$ ($\tilde X_c$)
rotate the bosons $\tilde x$ ($\tilde x^c$) from the flavour
to the mass states. Once SU(2)$_L$ is spontaneously broken, there
is also in general a nonzero mixing between the bosonic states
$\tilde x$ and $(\tilde x^c)^*$: their relative
importance is proportional to $m_W/m_{\tilde f}$,
which is, for our choice of the squark and
slepton masses, not bigger than $1/10$. We assume this to be
small enough to consider it as a perturbation.

The calculation itself is tedious but straightforward, and thus
we leave the details for the Appendix. We simply turn to the
systematic analysis of the possible solutions which keep proton
stable enough.

\section{Why proton decay does not rule out minimal SU(5)}

In this central section of our paper (the only one you should
read if you just wish to get our main point) we stick to the very
minimal SUSY SU(5) theory. In other words we assume
the conditions discussed above (valid when $M_{Pl}\to\infty$)
in the theory with only $5$ and $\bar 5$ light Higgses:

\begin{eqnarray}
\label{ab}
\underline A&=&\underline B=Y_U=Y_U^T\;,\\
\label{cd}
\underline C&=&\underline D=Y_D=Y_E\;,
\end{eqnarray}

\noindent
where, of course, these conditions are valid at the unification scale.
A quick glance at the Appendix shows that the longevity of the
proton can be achieved by, say, the following conditions at 1 GeV:

\begin{eqnarray}
\label{u}
(\tilde U^\dagger D)_{31,32}&\approx &0\;
,\\
(\tilde D^\dagger D)_{31,32}&\approx &0\;,\\
\label{uc}
(\tilde U_c^T Y_U^TD)_{31,32}&\approx &0\;,\\
\label{nena}
(\tilde N^T\underline C^T D)_{31,32}
(\tilde U^T\underline A D)_{32,31}&\approx &0\;,\\
(\tilde E_c^\dagger E_c)_{31,32}&\approx &0\;,\\
(\tilde D_c^\dagger D_c)_{31,32}&\approx &0\;,\\
(\tilde E^\dagger E)_{31,32}&\approx &0\;,\\
\label{ndva}
(\tilde N^\dagger E)_{31,32}&\approx &0\;.
\end{eqnarray}

If one wishes to quantify these conditions, one can not take
(\ref{ab})-(\ref{cd}) at face value, but instead must compute
the departure due to the running from $M_{GUT}$ to 1 GeV. It makes
no sense to do this here; after all, this is just a prototype example
and it can surely be satisfied at any scale. 

In the above equations we simply mean
that all the terms must be small. How small? It is
hard to quantify this precisely and, honestly speaking, it seems to
us a premature task. Our aim was to demonstrate that the theory is
still consistent with data and from the above formulae it is obvious.
If (when) proton decay is discovered and the decay modes
measured, it may be sensible to see how small should the above
terms be. Suffice it to say, that a percent suppression of the
super KM results should be enough \cite{Murayama:2001ur}. 
This means that on the average each vertex should be suppressed by
a factor of $1/3$ or so with respect to the minimal supergravity
predictions. It is very difficult to say more: in fact one could be
tempted to estimate that for example the combinations on the
lefthandsides of the above equations need to be at least $10^{-2}$ the same
combinations in super KM. However this is not authomatically neither
necessary nor enough. The fact is, that we have to do with a nonlinear  
system, since the total decay in a specified mode is proportional
to the square of a sum of single diagrams, each of them is proportional
to the product of four unknown mixings. Some of these mixings contribute to
different diagrams, and some depend on others, so the task of constraining
them numerically seem exagerate in view of our complete ignorance of all
these parameters. What we can say for sure is that if each of the diagrams
in the appendix is suppressed by a factor of $1/100$ with respect to
the minimal supergravity predictions, proton decay is not too fast and
minimal supersymmetric SU(5) is not ruled out.

Notice that all the terms can be made to vanish by a judicious choice
of squark and slepton mixing matrices. In other words, at this point
the proton decay limits provide information on the properties of
sfermions and {\it not} on the structure of the unified theory.

Notice further that the so called super KM basis, in which the
mixing angles of fermions and sfermions are equal, does not work
for the proton decay, since eqs. (\ref{u}), (\ref{uc}), (\ref{nena}) 
and (\ref{ndva}) are not satisfied. If you believe in super KM, 
you would conclude that the theory is ruled out. It is obvious 
though, from our work, that this is not true in general.

Notice even further, that all the relations (\ref{u})-(\ref{ndva})
do not require the extreme minimality conditions (\ref{ab})-(\ref{cd}).
More precisely, one can opt for the improvement of the fermion mass
relations and still save the proton.

One could worry that the above constraints for the sfermion
and fermion mixing matrices could be in contradiction with
the experimental bounds on the flavour violation low energy
processes. Fortunately, this is not true. Namely, the same
conditions (\ref{u})-(\ref{ndva}) suffice to render neutral
current flavour violation in-offensive (of course, the decoupling
is necessary for this to be true).

The analysis in the Appendix has been done with the assumption
of no left-right sfermion, neutralino or chargino mixing. As we
explained at the end of the previous section, this mixing can be
included in a perturbative way: one can show that, up to two
mass insertions, the same constraints (\ref{u})-(\ref{ndva})
kill all the contributions to nucleon decay. This is enough to
increase the nucleon lifetime above the experimental limit,
since each mixing multiplies the diagram by at least $1/10$.

Up to now we have discussed only the d=5 nucleon decay. What about
a generic d=6 contribution of gauge bosons relevant for both
ordinary and SUSY GUTs? In the very minimal case, $Y_D=Y_E$
and $Y_U=Y_U^T$, this is completely determined by the CKM
matrix \cite{Mohapatra:yj}. However, as soon one abandons this
unrealistic situation, this is not true anymore and the individual
up and down quark and lepton mixings enter the game and proton
decay is not as determined as before
\cite{Nandi:1982ew,Berezinsky:1983va}.

\section{Summary and outlook}

We hope to have convinced the reader that the supersymmetric SU(5)
theory even in its very minimal version is still alive and still
in accord with the nucleon decay limits. All that is required is
simply small mixing angles among squarks (sleptons) and/or quarks
(leptons), on top of the decoupling hypothesis, which sees the
first two generations of sfermions pushed to the 10 TeV region.

Does this mean that the proton decay experiments really probe
the sfermion and fermion mixing matrices? More precisely, are
there any other uncertainties involved in this game? At first
glance the answer is no. After all we have carefully defined the
minimal theory and found the predictions discussed above.
However, two points can still be raised.

(i) {\it Triplet-octet splitting (higher dimensional operators
in the Higgs sector).}

\noindent
In order to appreciate this point let us discuss the origin of
the problem in question. If one assumes that the heavy particles in
the adjoint superfield $\Sigma$ (the color octet and the weak triplet)
have the masses equal to $M_{GUT}$, the gauge couplings unify at
$\approx M_{GUT}\approx 10^{16}$ GeV. In this case the masses of
heavy triplets $T$ and $\bar T$ are smaller than $\approx
3.6\times 10^{15}$ GeV \cite{Murayama:2001ur}. A factor of around
$20$ increase of triplet masses according to \cite{Murayama:2001ur}
is sufficient to satisfy all the experimental constraints.

A simple possibility, which allows this, is to increase $M_{GUT}$
itself by a similar factor of $20$ or so. This turns out to be
easily satisfied by simply splitting the octet and triplet masses
in $\Sigma$ and allowing them to be smaller than $M_{GUT}$
\cite{Bachas:1995yt}.

Imagine for example that the octets and triplets are light
enough, so that their masses originate from dimension $4$
Planck scale induced terms in the superpotential, i.e. assume
that the renormalizable cubic term in the superpotential
(\ref{wh}) is negligible. In that case $m_3=4m_8$ 
\cite{Chkareuli:1998wi}, which at the one-loop level increases 
the proton decay mediating Higgs triplet masses by about a factor 
of $30$. 

(ii) {\it Improving the Yukawa sector with higher dimensional
operators.}

\noindent
In the minimal SU(5) theory and in the limit $M_{Pl}\to\infty$
the proton decay mediating Higgs triplet couplings are set by
SU(5) symmetry, since they must be equal to the ordinary doublet
couplings (\ref{ab})-(\ref{cd}). These relations can be, in the 
spirit of \cite{Dvali:1992hc}, changed by the nonrenormalizable 
$1/M_{Pl}$ suppressed operators 
\cite{Nath:1996qs,Nath:ft,Berezhiani:1998hg}.
This induces unfortunately additional uncertainty
in the constraints for the sfermion and fermion mixings.

In other words, to us
the nucleon decay not only can not rule out the structure of the
theory, but even in the case of observation would not easily
provide enough information about sfermion and fermion individual
mixings. In any case we see no reason whatsoever why one should
search for modifications of the theory at the GUT scale or
below for the sake of proton decay. If you need to do model
building, do not look here for an excuse.

\acknowledgments

We are extremely grateful to Ilia Gogoladze for
collaborating with us in the early stage of this work and
for numerous discussions and encouragement.
We thank Gia Dvali for discussions and comments,
Lotfi Boubeckeur and Jure Zupan for 
useful suggestions and Ariel Garcia
for discussions. G.S. is extremely grateful to
Alexei Smirnov for strong (although indirect) encouragement 
to publish these results. 
The work of B.B. is supported by the Ministry of Education, Science 
and Sport of the Republic of Slovenia; the work of G.S. is 
partially supported  by EEC under the TMR contracts ERBFMRX-CT960090 
and HPRN-CT-2000-00152. Both B.B. and P.F.P thank ICTP for hospitality 
during the course of this work.

\section*{Appendix}

In this Appendix we present the complete set of diagrams
responsible for d=5 nucleon decay in the minimal
supersymmetric SU(5) theory. In our notation $T$ and $\bar T$
stand for heavy Higgs triplets; $\tilde T$ and $\tilde{\bar T}$
denote their fermionic partners; $\tilde w^{\pm}$ stands for winos,
$\tilde h_{+,0}$ and $\tilde{\bar h}_{-,0}$ are light Higgsinos
and $\tilde V_0$ stand for neutral gauginos.

\vskip 0.5cm
\noindent
1)
$\underline{p\to (K^+,\pi^+,\rho^+,K^{*+}) \bar\nu_i}$,
$\underline{n\to (\pi^0,\rho^0,\eta,\omega,K^0,K^{*0})\bar\nu_i}$,
$\;\;$($i=1,2,3$)

\begin{equation}
\label{nw1}
\def\trgor{$\tilde T$}
\def\trdol{$\tilde{\bar T}$}
\def\sfgor{$\tilde t$}
\def\sfdol{$\tilde\tau$}
\def\spgor{$\tilde w^+$}
\def\spdol{$\tilde w^-$}
\def\iena{$d_{1,2}$}
\def\idva{$u$}
\def\itri{$\nu_i$}
\def\isti{$d_{2,1}$}
\begin{picture}(80,40)(40,20)
\SetWidth{0.8}
\ArrowLine(0,40)(20,40)
\ArrowLine(0,0)(20,0)
\ArrowLine(80,40)(60,40)
\ArrowLine(80,0)(60,0)
\DashLine(20,40)(60,40)3
\DashLine(20,0)(60,0)3
\ArrowLine(20,20)(20,40)
\ArrowLine(20,20)(20,0)
\Line(17,17)(23,23)
\Line(17,23)(23,17)
\ArrowLine(60,20)(60,40)
\ArrowLine(60,20)(60,0)
\Line(57,17)(63,23)
\Line(57,23)(63,17)
\Text(10,30)[]{\trgor}
\Text(10,10)[]{\trdol}
\Text(40,30)[]{\sfgor}
\Text(40,10)[]{\sfdol}
\Text(75,30)[]{\spgor}
\Text(75,10)[]{\spdol}
\Text(-10,40)[]{\iena}
\Text(-10,0)[]{\idva}
\Text(90,0)[]{\itri}
\Text(90,40)[]{\isti}
\end{picture}

\propto\;\;\;\;\;
(D^T\underline A\tilde U)_{13,23}(\tilde U^\dagger D)_{32,31}
(N^T\tilde E^*)_{i3}(\tilde E^T\underline C^TU)_{31}
\end{equation}

\begin{equation}
\def\tr{$T$}
\def\sfgor{$\tilde t$}
\def\sfdol{$\tilde\tau$}
\def\spgor{$\tilde w^+$}
\def\spdol{$\tilde w^-$}
\def\iena{$d_{1,2}$}
\def\idva{$u$}
\def\itri{$\nu_i$}
\def\isti{$d_{2,1}$}
\begin{picture}(80,40)(40,20)
\SetWidth{0.8}
\ArrowLine(0,40)(10,20)
\ArrowLine(0,0)(10,20)
\ArrowLine(80,40)(60,40)
\ArrowLine(80,0)(60,0)
\DashLine(10,20)(40,20)3
\DashLine(40,20)(60,40)3
\DashLine(40,20)(60,0)3
\ArrowLine(60,20)(60,40)
\ArrowLine(60,20)(60,0)
\Line(57,17)(63,23)
\Line(57,23)(63,17)
\Text(25,30)[]{\tr}
\Text(45,35)[]{\sfgor}
\Text(45,5)[]{\sfdol}
\Text(75,30)[]{\spgor}
\Text(75,10)[]{\spdol}
\Text(-10,40)[]{\iena}
\Text(-10,0)[]{\idva}
\Text(90,0)[]{\itri}
\Text(90,40)[]{\isti}
\end{picture}

\propto\;\;\;\;\;
(D^T\underline AU)_{11,21}(N^T\tilde E^*)_{i3}
(\tilde E^T\underline C^T\tilde U)_{33}(\tilde U^\dagger D)_{32,31}
\end{equation}

\begin{equation}
\def\trgor{$\tilde T$}
\def\trdol{$\tilde{\bar T}$}
\def\sfgor{$\tilde t$}
\def\sfdol{$\tilde b$}
\def\spgor{$\tilde w^+$}
\def\spdol{$\tilde w^-$}
\def\iena{$d_{1,2}$}
\def\idva{$\nu_i$}
\def\itri{$u$}
\def\isti{$d_{2,1}$}
\begin{picture}(80,40)(40,20)
\SetWidth{0.8}
\ArrowLine(0,40)(20,40)
\ArrowLine(0,0)(20,0)
\ArrowLine(80,40)(60,40)
\ArrowLine(80,0)(60,0)
\DashLine(20,40)(60,40)3
\DashLine(20,0)(60,0)3
\ArrowLine(20,20)(20,40)
\ArrowLine(20,20)(20,0)
\Line(17,17)(23,23)
\Line(17,23)(23,17)
\ArrowLine(60,20)(60,40)
\ArrowLine(60,20)(60,0)
\Line(57,17)(63,23)
\Line(57,23)(63,17)
\Text(10,30)[]{\trgor}
\Text(10,10)[]{\trdol}
\Text(40,30)[]{\sfgor}
\Text(40,10)[]{\sfdol}
\Text(75,30)[]{\spgor}
\Text(75,10)[]{\spdol}
\Text(-10,40)[]{\iena}
\Text(-10,0)[]{\idva}
\Text(90,0)[]{\itri}
\Text(90,40)[]{\isti}
\end{picture}

\propto\;\;\;\;\;
(D^T\underline A\tilde U)_{13,23}(\tilde U^\dagger D)_{32,31}
(U^T\tilde D^*)_{13}(\tilde D^T\underline CN)_{3i}
\end{equation}

\begin{equation}
\def\tr{$\bar T$}
\def\sfgor{$\tilde t$}
\def\sfdol{$\tilde b$}
\def\spgor{$\tilde w^+$}
\def\spdol{$\tilde w^-$}
\def\iena{$d_{1,2}$}
\def\idva{$\nu_i$}
\def\itri{$u$}
\def\isti{$d_{2,1}$}
\begin{picture}(80,40)(40,20)
\SetWidth{0.8}
\ArrowLine(0,40)(10,20)
\ArrowLine(0,0)(10,20)
\ArrowLine(80,40)(60,40)
\ArrowLine(80,0)(60,0)
\DashLine(10,20)(40,20)3
\DashLine(40,20)(60,40)3
\DashLine(40,20)(60,0)3
\ArrowLine(60,20)(60,40)
\ArrowLine(60,20)(60,0)
\Line(57,17)(63,23)
\Line(57,23)(63,17)
\Text(25,30)[]{\tr}
\Text(45,35)[]{\sfgor}
\Text(45,5)[]{\sfdol}
\Text(75,30)[]{\spgor}
\Text(75,10)[]{\spdol}
\Text(-10,40)[]{\iena}
\Text(-10,0)[]{\idva}
\Text(90,0)[]{\itri}
\Text(90,40)[]{\isti}
\end{picture}

\propto\;\;\;\;\;
(D^T\underline CN)_{1i,2i}(U^T\tilde D^*)_{13}
(\tilde D^T\underline A\tilde U)_{33}(\tilde U^\dagger D)_{32,31}
\end{equation}

\begin{equation}
\def\trgor{$\tilde T$}
\def\trdol{$\tilde{\bar T}$}
\def\sfgor{$\tilde t$}
\def\sfdol{$\tilde b$}
\def\spgor{$\tilde{\bar h}_-^\dagger$}
\def\spdol{$\tilde h_+^\dagger$}
\def\iena{$d_{1,2}$}
\def\idva{$\nu_i$}
\def\itri{${\bar u}^c$}
\def\isti{${\bar d}_{2,1}^c$}
\begin{picture}(80,40)(40,20)
\SetWidth{0.8}
\ArrowLine(0,40)(20,40)
\ArrowLine(0,0)(20,0)
\ArrowLine(80,40)(60,40)
\ArrowLine(80,0)(60,0)
\DashLine(20,40)(60,40)3
\DashLine(20,0)(60,0)3
\ArrowLine(20,20)(20,40)
\ArrowLine(20,20)(20,0)
\Line(17,17)(23,23)
\Line(17,23)(23,17)
\ArrowLine(60,20)(60,40)
\ArrowLine(60,20)(60,0)
\Line(57,17)(63,23)
\Line(57,23)(63,17)
\Text(10,30)[]{\trgor}
\Text(10,10)[]{\trdol}
\Text(40,30)[]{\sfgor}
\Text(40,10)[]{\sfdol}
\Text(75,30)[]{\spgor}
\Text(75,10)[]{\spdol}
\Text(-10,40)[]{\iena}
\Text(-10,0)[]{\idva}
\Text(90,0)[]{\itri}
\Text(90,40)[]{\isti}
\end{picture}

\propto\;\;\;\;\;
(D^T\underline A\tilde U)_{13,23}(\tilde U^\dagger Y_D^*D_c^*)_{32,31}
(U_c^\dagger Y_U^\dagger\tilde D^*)_{13}(\tilde D^T\underline CN)_{3i}
\end{equation}

\begin{equation}
\label{nh2}
\def\tr{$\bar T$}
\def\sfgor{$\tilde t$}
\def\sfdol{$\tilde b$}
\def\spgor{$\tilde{\bar h}_-^\dagger$}
\def\spdol{$\tilde h_+^\dagger$}
\def\iena{$d_{1,2}$}
\def\idva{$\nu_i$}
\def\itri{${\bar u}^c$}
\def\isti{${\bar d}_{2,1}^c$}
\begin{picture}(80,40)(40,20)
\SetWidth{0.8}
\ArrowLine(0,40)(10,20)
\ArrowLine(0,0)(10,20)
\ArrowLine(80,40)(60,40)
\ArrowLine(80,0)(60,0)
\DashLine(10,20)(40,20)3
\DashLine(40,20)(60,40)3
\DashLine(40,20)(60,0)3
\ArrowLine(60,20)(60,40)
\ArrowLine(60,20)(60,0)
\Line(57,17)(63,23)
\Line(57,23)(63,17)
\Text(25,30)[]{\tr}
\Text(45,35)[]{\sfgor}
\Text(45,5)[]{\sfdol}
\Text(75,30)[]{\spgor}
\Text(75,10)[]{\spdol}
\Text(-10,40)[]{\iena}
\Text(-10,0)[]{\idva}
\Text(90,0)[]{\itri}
\Text(90,40)[]{\isti}
\end{picture}

\propto\;\;\;\;\;
(D^T\underline CN)_{1i,2i}(U_c^\dagger Y_U^\dagger\tilde D^*)_{13}
(\tilde D^T\underline A\tilde U)_{33}(\tilde U^\dagger Y_D^*D_c^*)_{32,31}
\end{equation}

\begin{equation}
\label{nh3}
\def\trgor{$\tilde T^\dagger$}
\def\trdol{$\tilde{\bar T}^\dagger$}
\def\sfgor{$\tilde\tau^c$}
\def\sfdol{$\tilde t^c$}
\def\spgor{$\tilde{\bar h}_-$}
\def\spdol{$\tilde h_+$}
\def\iena{${\bar u}^c$}
\def\idva{${\bar d}_{2,1}^c$}
\def\itri{$d_{1,2}$}
\def\isti{$\nu_i$}
\begin{picture}(80,40)(40,20)
\SetWidth{0.8}
\ArrowLine(0,40)(20,40)
\ArrowLine(0,0)(20,0)
\ArrowLine(80,40)(60,40)
\ArrowLine(80,0)(60,0)
\DashLine(20,40)(60,40)3
\DashLine(20,0)(60,0)3
\ArrowLine(20,20)(20,40)
\ArrowLine(20,20)(20,0)
\Line(17,17)(23,23)
\Line(17,23)(23,17)
\ArrowLine(60,20)(60,40)
\ArrowLine(60,20)(60,0)
\Line(57,17)(63,23)
\Line(57,23)(63,17)
\Text(10,30)[]{\trgor}
\Text(10,10)[]{\trdol}
\Text(40,30)[]{\sfgor}
\Text(40,10)[]{\sfdol}
\Text(75,30)[]{\spgor}
\Text(75,10)[]{\spdol}
\Text(-10,40)[]{\iena}
\Text(-10,0)[]{\idva}
\Text(90,0)[]{\itri}
\Text(90,40)[]{\isti}
\end{picture}

\propto\;\;\;\;\;
(U_c^\dagger\underline B^*\tilde E_c^*)_{13}(\tilde E_c^TY_EN)_{3i}
(D^TY_U\tilde U_c)_{13,23}(\tilde U_c^\dagger\underline D^*D_c^*)_{32,31}
\end{equation}

\begin{equation}
\label{nh4}
\def\tr{$\bar T$}
\def\sfgor{$\tilde\tau^c$}
\def\sfdol{$\tilde t^c$}
\def\spgor{$\tilde{\bar h}_-$}
\def\spdol{$\tilde h_+$}
\def\iena{${\bar u}^c$}
\def\idva{${\bar d}_{1,2}^c$}
\def\itri{$d_{2,1}$}
\def\isti{$\nu_i$}
\begin{picture}(80,40)(40,20)
\SetWidth{0.8}
\ArrowLine(0,40)(10,20)
\ArrowLine(0,0)(10,20)
\ArrowLine(80,40)(60,40)
\ArrowLine(80,0)(60,0)
\DashLine(10,20)(40,20)3
\DashLine(40,20)(60,40)3
\DashLine(40,20)(60,0)3
\ArrowLine(60,20)(60,40)
\ArrowLine(60,20)(60,0)
\Line(57,17)(63,23)
\Line(57,23)(63,17)
\Text(25,30)[]{\tr}
\Text(45,35)[]{\sfgor}
\Text(45,5)[]{\sfdol}
\Text(75,30)[]{\spgor}
\Text(75,10)[]{\spdol}
\Text(-10,40)[]{\iena}
\Text(-10,0)[]{\idva}
\Text(90,0)[]{\itri}
\Text(90,40)[]{\isti}
\end{picture}

\propto\;\;\;\;\;
(U_c^\dagger\underline D^* D_c^*)_{11,12}(D^TY_U\tilde U_c)_{23,13}
(\tilde U_c^\dagger\underline B^*\tilde E_c^*)_{33}(\tilde E_c^TY_EN)_{3i}
\end{equation}

\begin{equation}
\label{nh01}
\def\trgor{$\tilde T$}
\def\trdol{$\tilde{\bar T}$}
\def\sfgor{$\tilde t$}
\def\sfdol{$\tilde b$}
\def\spgor{$\tilde h_0^\dagger$}
\def\spdol{$\tilde{\bar h}_0^\dagger$}
\def\iena{$d_{1,2}$}
\def\idva{$\nu_i$}
\def\itri{${\bar d}_{2,1}^c$}
\def\isti{${\bar u}^c$}
\begin{picture}(80,40)(40,20)
\SetWidth{0.8}
\ArrowLine(0,40)(20,40)
\ArrowLine(0,0)(20,0)
\ArrowLine(80,40)(60,40)
\ArrowLine(80,0)(60,0)
\DashLine(20,40)(60,40)3
\DashLine(20,0)(60,0)3
\ArrowLine(20,20)(20,40)
\ArrowLine(20,20)(20,0)
\Line(17,17)(23,23)
\Line(17,23)(23,17)
\ArrowLine(60,20)(60,40)
\ArrowLine(60,20)(60,0)
\Line(57,17)(63,23)
\Line(57,23)(63,17)
\Text(10,30)[]{\trgor}
\Text(10,10)[]{\trdol}
\Text(40,30)[]{\sfgor}
\Text(40,10)[]{\sfdol}
\Text(75,30)[]{\spgor}
\Text(75,10)[]{\spdol}
\Text(-10,40)[]{\iena}
\Text(-10,0)[]{\idva}
\Text(90,0)[]{\itri}
\Text(90,40)[]{\isti}
\end{picture}

\propto\;\;\;\;\;
(D^T\underline A\tilde U)_{13,23}(\tilde U^\dagger Y_U^*U_c^*)_{31}
(D_c^\dagger Y_D^\dagger\tilde D^*)_{23,13}(\tilde D^T\underline CN)_{3i}
\end{equation}

\begin{equation}
\def\tr{$\bar T$}
\def\sfgor{$\tilde b$}
\def\sfdol{$\tilde t$}
\def\spgor{$\tilde{\bar h}_0^\dagger$}
\def\spdol{$\tilde h_0^\dagger$}
\def\iena{$d_{1,2}$}
\def\idva{$\nu_i$}
\def\itri{${\bar u}^c$}
\def\isti{${\bar d}_{2,1}^c$}
\begin{picture}(80,40)(40,20)
\SetWidth{0.8}
\ArrowLine(0,40)(10,20)
\ArrowLine(0,0)(10,20)
\ArrowLine(80,40)(60,40)
\ArrowLine(80,0)(60,0)
\DashLine(10,20)(40,20)3
\DashLine(40,20)(60,40)3
\DashLine(40,20)(60,0)3
\ArrowLine(60,20)(60,40)
\ArrowLine(60,20)(60,0)
\Line(57,17)(63,23)
\Line(57,23)(63,17)
\Text(25,30)[]{\tr}
\Text(45,35)[]{\sfgor}
\Text(45,5)[]{\sfdol}
\Text(75,30)[]{\spgor}
\Text(75,10)[]{\spdol}
\Text(-10,40)[]{\iena}
\Text(-10,0)[]{\idva}
\Text(90,0)[]{\itri}
\Text(90,40)[]{\isti}
\end{picture}

\propto\;\;\;\;\;
(D^T\underline CN)_{1i,2i}(U_c^\dagger Y_U^\dagger\tilde U^*)_{13}
(\tilde U^T\underline A\tilde D)_{33}(\tilde D^\dagger Y_D^*D_c^*)_{32,31}
\end{equation}

\begin{equation}
\def\trgor{$\tilde T$}
\def\trdol{$\tilde{\bar T}$}
\def\sfgor{$\tilde t$}
\def\sfdol{$\tilde b$}
\def\spgor{$\tilde V_0$}
\def\spdol{$\tilde V_0$}
\def\iena{$d_{1,2}$}
\def\idva{$\nu_i$}
\def\itri{$d_{2,1}$}
\def\isti{$u$}
\begin{picture}(80,40)(40,20)
\SetWidth{0.8}
\ArrowLine(0,40)(20,40)
\ArrowLine(0,0)(20,0)
\ArrowLine(80,40)(60,40)
\ArrowLine(80,0)(60,0)
\DashLine(20,40)(60,40)3
\DashLine(20,0)(60,0)3
\ArrowLine(20,20)(20,40)
\ArrowLine(20,20)(20,0)
\Line(17,17)(23,23)
\Line(17,23)(23,17)
\ArrowLine(60,20)(60,40)
\ArrowLine(60,20)(60,0)
\Line(57,17)(63,23)
\Line(57,23)(63,17)
\Text(10,30)[]{\trgor}
\Text(10,10)[]{\trdol}
\Text(40,30)[]{\sfgor}
\Text(40,10)[]{\sfdol}
\Text(75,30)[]{\spgor}
\Text(75,10)[]{\spdol}
\Text(-10,40)[]{\iena}
\Text(-10,0)[]{\idva}
\Text(90,0)[]{\itri}
\Text(90,40)[]{\isti}
\end{picture}

\propto\;\;\;\;\;
(D^T\underline A\tilde U)_{13,23}(\tilde U^\dagger U)_{31}
(D^T\tilde D^*)_{23,13}(\tilde D^T\underline CN)_{3i}
\end{equation}

\begin{equation}
\def\tr{$\bar T$}
\def\sfgor{$\tilde t$}
\def\sfdol{$\tilde b$}
\def\spgor{$\tilde V_0$}
\def\spdol{$\tilde V_0$}
\def\iena{$d_{1,2}$}
\def\idva{$\nu_i$}
\def\itri{$d_{2,1}$}
\def\isti{$u$}
\begin{picture}(80,40)(40,20)
\SetWidth{0.8}
\ArrowLine(0,40)(10,20)
\ArrowLine(0,0)(10,20)
\ArrowLine(80,40)(60,40)
\ArrowLine(80,0)(60,0)
\DashLine(10,20)(40,20)3
\DashLine(40,20)(60,40)3
\DashLine(40,20)(60,0)3
\ArrowLine(60,20)(60,40)
\ArrowLine(60,20)(60,0)
\Line(57,17)(63,23)
\Line(57,23)(63,17)
\Text(25,30)[]{\tr}
\Text(45,35)[]{\sfgor}
\Text(45,5)[]{\sfdol}
\Text(75,30)[]{\spgor}
\Text(75,10)[]{\spdol}
\Text(-10,40)[]{\iena}
\Text(-10,0)[]{\idva}
\Text(90,0)[]{\itri}
\Text(90,40)[]{\isti}
\end{picture}

\propto\;\;\;\;\;
(D^T\underline CN)_{1i,2i}(D^T\tilde D^*)_{23,13}
(\tilde D^T\underline A\tilde U)_{33}(\tilde U^\dagger U)_{31}
\end{equation}

\begin{equation}
\def\trgor{$\tilde{\bar T}$}
\def\trdol{$\tilde T$}
\def\sfgor{$\tilde\nu$}
\def\sfdol{$\tilde b$}
\def\spgor{$\tilde V_0$}
\def\spdol{$\tilde V_0$}
\def\iena{$d_{1,2}$}
\def\idva{$u$}
\def\itri{$d_{2,1}$}
\def\isti{$\nu_i$}
\begin{picture}(80,40)(40,20)
\SetWidth{0.8}
\ArrowLine(0,40)(20,40)
\ArrowLine(0,0)(20,0)
\ArrowLine(80,40)(60,40)
\ArrowLine(80,0)(60,0)
\DashLine(20,40)(60,40)3
\DashLine(20,0)(60,0)3
\ArrowLine(20,20)(20,40)
\ArrowLine(20,20)(20,0)
\Line(17,17)(23,23)
\Line(17,23)(23,17)
\ArrowLine(60,20)(60,40)
\ArrowLine(60,20)(60,0)
\Line(57,17)(63,23)
\Line(57,23)(63,17)
\Text(10,30)[]{\trgor}
\Text(10,10)[]{\trdol}
\Text(40,30)[]{\sfgor}
\Text(40,10)[]{\sfdol}
\Text(75,30)[]{\spgor}
\Text(75,10)[]{\spdol}
\Text(-10,40)[]{\iena}
\Text(-10,0)[]{\idva}
\Text(90,0)[]{\itri}
\Text(90,40)[]{\isti}
\end{picture}

\propto\;\;\;\;\;
(D^T\underline C\tilde N)_{13,23}(\tilde N^\dagger N)_{3i}
(D^T\tilde D^*)_{23,13}(\tilde D^T\underline AU)_{31}
\end{equation}

\begin{equation}
\label{nv4}
\def\tr{$T$}
\def\sfgor{$\tilde\nu$}
\def\sfdol{$\tilde b$}
\def\spgor{$\tilde V_0$}
\def\spdol{$\tilde V_0$}
\def\iena{$u$}
\def\idva{$d_{1,2}$}
\def\itri{$d_{2,1}$}
\def\isti{$\nu_i$}
\begin{picture}(80,40)(40,20)
\SetWidth{0.8}
\ArrowLine(0,40)(10,20)
\ArrowLine(0,0)(10,20)
\ArrowLine(80,40)(60,40)
\ArrowLine(80,0)(60,0)
\DashLine(10,20)(40,20)3
\DashLine(40,20)(60,40)3
\DashLine(40,20)(60,0)3
\ArrowLine(60,20)(60,40)
\ArrowLine(60,20)(60,0)
\Line(57,17)(63,23)
\Line(57,23)(63,17)
\Text(25,30)[]{\tr}
\Text(45,35)[]{\sfgor}
\Text(45,5)[]{\sfdol}
\Text(75,30)[]{\spgor}
\Text(75,10)[]{\spdol}
\Text(-10,40)[]{\iena}
\Text(-10,0)[]{\idva}
\Text(90,0)[]{\itri}
\Text(90,40)[]{\isti}
\end{picture}

\propto\;\;\;\;\;
(U^T\underline AD)_{11,12}(D^T\tilde D^*)_{23,13}
(\tilde D^T\underline C\tilde N)_{33}(\tilde N^\dagger N)_{3i}
\end{equation}

\begin{equation}
\label{nv5}
\def\trgor{$\tilde{\bar T}$}
\def\trdol{$\tilde T$}
\def\sfgor{$\tilde\nu$}
\def\sfdol{$\tilde t$}
\def\spgor{$\tilde V_0$}
\def\spdol{$\tilde V_0$}
\def\iena{$d_{1,2}$}
\def\idva{$d_{2,1}$}
\def\itri{$u$}
\def\isti{$\nu_i$}
\begin{picture}(80,40)(40,20)
\SetWidth{0.8}
\ArrowLine(0,40)(20,40)
\ArrowLine(0,0)(20,0)
\ArrowLine(80,40)(60,40)
\ArrowLine(80,0)(60,0)
\DashLine(20,40)(60,40)3
\DashLine(20,0)(60,0)3
\ArrowLine(20,20)(20,40)
\ArrowLine(20,20)(20,0)
\Line(17,17)(23,23)
\Line(17,23)(23,17)
\ArrowLine(60,20)(60,40)
\ArrowLine(60,20)(60,0)
\Line(57,17)(63,23)
\Line(57,23)(63,17)
\Text(10,30)[]{\trgor}
\Text(10,10)[]{\trdol}
\Text(40,30)[]{\sfgor}
\Text(40,10)[]{\sfdol}
\Text(75,30)[]{\spgor}
\Text(75,10)[]{\spdol}
\Text(-10,40)[]{\iena}
\Text(-10,0)[]{\idva}
\Text(90,0)[]{\itri}
\Text(90,40)[]{\isti}
\end{picture}

\propto\;\;\;\;\;
(D^T\underline C\tilde N)_{13,23}(\tilde N^\dagger N)_{3i}
(U^T\tilde U^*)_{13}(\tilde U^T\underline AD)_{32,31}
\end{equation}

\vskip 1cm

\noindent
2) $\underline{p\to (K^0,\pi^0,\eta, K^{*0},\rho^0,\omega) e_i^+}$,
$\underline{n\to (K^-,\pi^-,K^{*-},\rho^-) e_i^+}$
$\;\;$($i=1,2$, for $K^*$ only $i=1$)

\begin{equation}
\def\trgor{$\tilde{\bar T}$}
\def\trdol{$\tilde T$}
\def\sfgor{$\tilde\nu$}
\def\sfdol{$\tilde b$}
\def\spgor{$\tilde w^+$}
\def\spdol{$\tilde w^-$}
\def\iena{$d_{1,2}$}
\def\idva{$u$}
\def\itri{$u$}
\def\isti{$e_i$}
\begin{picture}(80,40)(40,20)
\SetWidth{0.8}
\ArrowLine(0,40)(20,40)
\ArrowLine(0,0)(20,0)
\ArrowLine(80,40)(60,40)
\ArrowLine(80,0)(60,0)
\DashLine(20,40)(60,40)3
\DashLine(20,0)(60,0)3
\ArrowLine(20,20)(20,40)
\ArrowLine(20,20)(20,0)
\Line(17,17)(23,23)
\Line(17,23)(23,17)
\ArrowLine(60,20)(60,40)
\ArrowLine(60,20)(60,0)
\Line(57,17)(63,23)
\Line(57,23)(63,17)
\Text(10,30)[]{\trgor}
\Text(10,10)[]{\trdol}
\Text(40,30)[]{\sfgor}
\Text(40,10)[]{\sfdol}
\Text(75,30)[]{\spgor}
\Text(75,10)[]{\spdol}
\Text(-10,40)[]{\iena}
\Text(-10,0)[]{\idva}
\Text(90,0)[]{\itri}
\Text(90,40)[]{\isti}
\end{picture}

\propto\;\;\;\;\;
(D^T\underline C\tilde N)_{13,23}(\tilde N^\dagger E)_{3i}
(U^T\tilde D^*)_{13}(\tilde D^T\underline AU)_{31}
\end{equation}

\begin{equation}
\def\tr{$T$}
\def\sfgor{$\tilde\nu$}
\def\sfdol{$\tilde b$}
\def\spgor{$\tilde w^+$}
\def\spdol{$\tilde w^-$}
\def\iena{$d_{1,2}$}
\def\idva{$u$}
\def\itri{$u$}
\def\isti{$e_i$}
\begin{picture}(80,40)(40,20)
\SetWidth{0.8}
\ArrowLine(0,40)(10,20)
\ArrowLine(0,0)(10,20)
\ArrowLine(80,40)(60,40)
\ArrowLine(80,0)(60,0)
\DashLine(10,20)(40,20)3
\DashLine(40,20)(60,40)3
\DashLine(40,20)(60,0)3
\ArrowLine(60,20)(60,40)
\ArrowLine(60,20)(60,0)
\Line(57,17)(63,23)
\Line(57,23)(63,17)
\Text(25,30)[]{\tr}
\Text(45,35)[]{\sfgor}
\Text(45,5)[]{\sfdol}
\Text(75,30)[]{\spgor}
\Text(75,10)[]{\spdol}
\Text(-10,40)[]{\iena}
\Text(-10,0)[]{\idva}
\Text(90,0)[]{\itri}
\Text(90,40)[]{\isti}
\end{picture}

\propto\;\;\;\;\;
(D^T\underline AU)_{11,21}(U^T\tilde D^*)_{13}
(\tilde D^T\underline C\tilde N)_{33}(\tilde N^\dagger E)_{3i}
\end{equation}

\begin{equation}
\def\trgor{$\tilde T$}
\def\trdol{$\tilde{\bar T}$}
\def\sfgor{$\tilde b$}
\def\sfdol{$\tilde t$}
\def\spgor{$\tilde w^-$}
\def\spdol{$\tilde w^+$}
\def\iena{$u$}
\def\idva{$e_i$}
\def\itri{$d_{1,2}$}
\def\isti{$u$}
\begin{picture}(80,40)(40,20)
\SetWidth{0.8}
\ArrowLine(0,40)(20,40)
\ArrowLine(0,0)(20,0)
\ArrowLine(80,40)(60,40)
\ArrowLine(80,0)(60,0)
\DashLine(20,40)(60,40)3
\DashLine(20,0)(60,0)3
\ArrowLine(20,20)(20,40)
\ArrowLine(20,20)(20,0)
\Line(17,17)(23,23)
\Line(17,23)(23,17)
\ArrowLine(60,20)(60,40)
\ArrowLine(60,20)(60,0)
\Line(57,17)(63,23)
\Line(57,23)(63,17)
\Text(10,30)[]{\trgor}
\Text(10,10)[]{\trdol}
\Text(40,30)[]{\sfgor}
\Text(40,10)[]{\sfdol}
\Text(75,30)[]{\spgor}
\Text(75,10)[]{\spdol}
\Text(-10,40)[]{\iena}
\Text(-10,0)[]{\idva}
\Text(90,0)[]{\itri}
\Text(90,40)[]{\isti}
\end{picture}

\propto\;\;\;\;\;
(U^T\underline A\tilde D)_{13}(\tilde D^\dagger U)_{31}
(D^T\tilde U^*)_{13,23}(\tilde U^T\underline CE)_{3i}
\end{equation}

\begin{equation}
\def\tr{$\bar T$}
\def\sfgor{$\tilde b$}
\def\sfdol{$\tilde t$}
\def\spgor{$\tilde w^-$}
\def\spdol{$\tilde w^+$}
\def\iena{$u$}
\def\idva{$e_i$}
\def\itri{$d_{1,2}$}
\def\isti{$u$}
\begin{picture}(80,40)(40,20)
\SetWidth{0.8}
\ArrowLine(0,40)(10,20)
\ArrowLine(0,0)(10,20)
\ArrowLine(80,40)(60,40)
\ArrowLine(80,0)(60,0)
\DashLine(10,20)(40,20)3
\DashLine(40,20)(60,40)3
\DashLine(40,20)(60,0)3
\ArrowLine(60,20)(60,40)
\ArrowLine(60,20)(60,0)
\Line(57,17)(63,23)
\Line(57,23)(63,17)
\Text(25,30)[]{\tr}
\Text(45,35)[]{\sfgor}
\Text(45,5)[]{\sfdol}
\Text(75,30)[]{\spgor}
\Text(75,10)[]{\spdol}
\Text(-10,40)[]{\iena}
\Text(-10,0)[]{\idva}
\Text(90,0)[]{\itri}
\Text(90,40)[]{\isti}
\end{picture}

\propto\;\;\;\;\;
(U^T\underline CE)_{1i}(D^T\tilde U^*)_{13,23}
(\tilde U^T\underline A\tilde D)_{33}(\tilde D^\dagger U)_{31}
\end{equation}

\begin{equation}
\def\trgor{$\tilde{\bar T}^\dagger$}
\def\trdol{$\tilde T^\dagger$}
\def\sfgor{$\tilde t^c$}
\def\sfdol{$\tilde b^c$}
\def\spgor{$\tilde h_+$}
\def\spdol{$\tilde{\bar h}_-$}
\def\iena{${\bar e}_i^c$}
\def\idva{${\bar u}^c$}
\def\itri{$u$}
\def\isti{$d_{1,2}$}
\begin{picture}(80,40)(40,20)
\SetWidth{0.8}
\ArrowLine(0,40)(20,40)
\ArrowLine(0,0)(20,0)
\ArrowLine(80,40)(60,40)
\ArrowLine(80,0)(60,0)
\DashLine(20,40)(60,40)3
\DashLine(20,0)(60,0)3
\ArrowLine(20,20)(20,40)
\ArrowLine(20,20)(20,0)
\Line(17,17)(23,23)
\Line(17,23)(23,17)
\ArrowLine(60,20)(60,40)
\ArrowLine(60,20)(60,0)
\Line(57,17)(63,23)
\Line(57,23)(63,17)
\Text(10,30)[]{\trgor}
\Text(10,10)[]{\trdol}
\Text(40,30)[]{\sfgor}
\Text(40,10)[]{\sfdol}
\Text(75,30)[]{\spgor}
\Text(75,10)[]{\spdol}
\Text(-10,40)[]{\iena}
\Text(-10,0)[]{\idva}
\Text(90,0)[]{\itri}
\Text(90,40)[]{\isti}
\end{picture}

\propto\;\;\;\;\;
(E_c^\dagger\underline B^\dagger\tilde U_c^*)_{i3}
(\tilde U_c^TY_U^TD)_{31,32}(U^TY_D\tilde D_c)_{13}
(\tilde D_c^\dagger\underline D^\dagger U_c^*)_{31}
\end{equation}

\begin{equation}
\def\tr{$T$}
\def\sfgor{$\tilde t^c$}
\def\sfdol{$\tilde b^c$}
\def\spgor{$\tilde h_+$}
\def\spdol{$\tilde{\bar h}_-$}
\def\iena{${\bar e}_i^c$}
\def\idva{${\bar u}^c$}
\def\itri{$u$}
\def\isti{$d_{1,2}$}
\begin{picture}(80,40)(40,20)
\SetWidth{0.8}
\ArrowLine(0,40)(10,20)
\ArrowLine(0,0)(10,20)
\ArrowLine(80,40)(60,40)
\ArrowLine(80,0)(60,0)
\DashLine(10,20)(40,20)3
\DashLine(40,20)(60,40)3
\DashLine(40,20)(60,0)3
\ArrowLine(60,20)(60,40)
\ArrowLine(60,20)(60,0)
\Line(57,17)(63,23)
\Line(57,23)(63,17)
\Text(25,30)[]{\tr}
\Text(45,35)[]{\sfgor}
\Text(45,5)[]{\sfdol}
\Text(75,30)[]{\spgor}
\Text(75,10)[]{\spdol}
\Text(-10,40)[]{\iena}
\Text(-10,0)[]{\idva}
\Text(90,0)[]{\itri}
\Text(90,40)[]{\isti}
\end{picture}

\propto\;\;\;\;\;
(E_c^\dagger\underline B^\dagger U_c^*)_{i1}(U^TY_D\tilde D_c)_{13}
(\tilde D_c^\dagger\underline D^\dagger\tilde U_c^*)_{33}
(\tilde U_c^TY_U^TD)_{31,32}
\end{equation}

\begin{equation}
\def\trgor{$\tilde T$}
\def\trdol{$\tilde{\bar T}$}
\def\sfgor{$\tilde b$}
\def\sfdol{$\tilde t$}
\def\spgor{$\tilde h_+^\dagger$}
\def\spdol{$\tilde{\bar h}_-^\dagger$}
\def\iena{$u$}
\def\idva{$e_i$}
\def\itri{${\bar d}_{1,2}^c$}
\def\isti{${\bar u}^c$}
\begin{picture}(80,40)(40,20)
\SetWidth{0.8}
\ArrowLine(0,40)(20,40)
\ArrowLine(0,0)(20,0)
\ArrowLine(80,40)(60,40)
\ArrowLine(80,0)(60,0)
\DashLine(20,40)(60,40)3
\DashLine(20,0)(60,0)3
\ArrowLine(20,20)(20,40)
\ArrowLine(20,20)(20,0)
\Line(17,17)(23,23)
\Line(17,23)(23,17)
\ArrowLine(60,20)(60,40)
\ArrowLine(60,20)(60,0)
\Line(57,17)(63,23)
\Line(57,23)(63,17)
\Text(10,30)[]{\trgor}
\Text(10,10)[]{\trdol}
\Text(40,30)[]{\sfgor}
\Text(40,10)[]{\sfdol}
\Text(75,30)[]{\spgor}
\Text(75,10)[]{\spdol}
\Text(-10,40)[]{\iena}
\Text(-10,0)[]{\idva}
\Text(90,0)[]{\itri}
\Text(90,40)[]{\isti}
\end{picture}

\propto\;\;\;\;\;
(U^T\underline A\tilde D)_{13}(\tilde D^\dagger Y_U^*U_c^*)_{31}
(D_c^\dagger Y_D^\dagger\tilde U^*)_{13,23}(\tilde U^T\underline CE)_{3i}
\end{equation}

\begin{equation}
\def\tr{$\bar T$}
\def\sfgor{$\tilde b$}
\def\sfdol{$\tilde t$}
\def\spgor{$\tilde h_+^\dagger$}
\def\spdol{$\tilde{\bar h}_-^\dagger$}
\def\iena{$u$}
\def\idva{$e_i$}
\def\itri{${\bar d}_{1,2}^c$}
\def\isti{${\bar u}^c$}
\begin{picture}(80,40)(40,20)
\SetWidth{0.8}
\ArrowLine(0,40)(10,20)
\ArrowLine(0,0)(10,20)
\ArrowLine(80,40)(60,40)
\ArrowLine(80,0)(60,0)
\DashLine(10,20)(40,20)3
\DashLine(40,20)(60,40)3
\DashLine(40,20)(60,0)3
\ArrowLine(60,20)(60,40)
\ArrowLine(60,20)(60,0)
\Line(57,17)(63,23)
\Line(57,23)(63,17)
\Text(25,30)[]{\tr}
\Text(45,35)[]{\sfgor}
\Text(45,5)[]{\sfdol}
\Text(75,30)[]{\spgor}
\Text(75,10)[]{\spdol}
\Text(-10,40)[]{\iena}
\Text(-10,0)[]{\idva}
\Text(90,0)[]{\itri}
\Text(90,40)[]{\isti}
\end{picture}

\propto\;\;\;\;\;
(U^T\underline CE)_{1i}(D_c^\dagger Y_D^\dagger\tilde U^*)_{13,23}
(\tilde U^T\underline A\tilde D)_{33}(\tilde D^\dagger Y_U^*U_c^*)_{31}
\end{equation}

\begin{equation}
\def\trgor{$\tilde{\bar T}$}
\def\trdol{$\tilde T$}
\def\sfgor{$\tilde\nu$}
\def\sfdol{$\tilde b$}
\def\spgor{$\tilde{\bar h}_-^\dagger$}
\def\spdol{$\tilde h_+^\dagger$}
\def\iena{$d_{1,2}$}
\def\idva{$u$}
\def\itri{${\bar u}^c$}
\def\isti{${\bar e}_i^c$}
\begin{picture}(80,40)(40,20)
\SetWidth{0.8}
\ArrowLine(0,40)(20,40)
\ArrowLine(0,0)(20,0)
\ArrowLine(80,40)(60,40)
\ArrowLine(80,0)(60,0)
\DashLine(20,40)(60,40)3
\DashLine(20,0)(60,0)3
\ArrowLine(20,20)(20,40)
\ArrowLine(20,20)(20,0)
\Line(17,17)(23,23)
\Line(17,23)(23,17)
\ArrowLine(60,20)(60,40)
\ArrowLine(60,20)(60,0)
\Line(57,17)(63,23)
\Line(57,23)(63,17)
\Text(10,30)[]{\trgor}
\Text(10,10)[]{\trdol}
\Text(40,30)[]{\sfgor}
\Text(40,10)[]{\sfdol}
\Text(75,30)[]{\spgor}
\Text(75,10)[]{\spdol}
\Text(-10,40)[]{\iena}
\Text(-10,0)[]{\idva}
\Text(90,0)[]{\itri}
\Text(90,40)[]{\isti}
\end{picture}

\propto\;\;\;\;\;
(D^T\underline C\tilde N)_{13,23}(\tilde N^\dagger Y_E^\dagger E_c^*)_{3i}
(U_c^\dagger Y_U^\dagger\tilde D^*)_{13}(\tilde D^T\underline AU)_{31}
\end{equation}

\begin{equation}
\def\tr{$T$}
\def\sfgor{$\tilde\nu$}
\def\sfdol{$\tilde b$}
\def\spgor{$\tilde{\bar h}_-^\dagger$}
\def\spdol{$\tilde h_+^\dagger$}
\def\iena{$d_{1,2}$}
\def\idva{$u$}
\def\itri{${\bar u}^c$}
\def\isti{${\bar e}_i^c$}
\begin{picture}(80,40)(40,20)
\SetWidth{0.8}
\ArrowLine(0,40)(10,20)
\ArrowLine(0,0)(10,20)
\ArrowLine(80,40)(60,40)
\ArrowLine(80,0)(60,0)
\DashLine(10,20)(40,20)3
\DashLine(40,20)(60,40)3
\DashLine(40,20)(60,0)3
\ArrowLine(60,20)(60,40)
\ArrowLine(60,20)(60,0)
\Line(57,17)(63,23)
\Line(57,23)(63,17)
\Text(25,30)[]{\tr}
\Text(45,35)[]{\sfgor}
\Text(45,5)[]{\sfdol}
\Text(75,30)[]{\spgor}
\Text(75,10)[]{\spdol}
\Text(-10,40)[]{\iena}
\Text(-10,0)[]{\idva}
\Text(90,0)[]{\itri}
\Text(90,40)[]{\isti}
\end{picture}

\propto\;\;\;\;\;
(D^T\underline AU)_{11,21}(U_c^\dagger Y_U^\dagger\tilde
D^*)_{13}(\tilde D^T\underline C\tilde N)_{33}
(\tilde N^\dagger Y_E^\dagger E_c^*)_{3i}
\end{equation}

\begin{equation}
\def\trgor{$\tilde T$}
\def\trdol{$\tilde{\bar T}$}
\def\sfgor{$\tilde b$}
\def\sfdol{$\tilde t$}
\def\spgor{$\tilde{\bar h}_0^\dagger$}
\def\spdol{$\tilde h_0^\dagger$}
\def\iena{$u$}
\def\idva{$e_i$}
\def\itri{${\bar u}^c$}
\def\isti{${\bar d}_{1,2}^c$}
\begin{picture}(80,40)(40,20)
\SetWidth{0.8}
\ArrowLine(0,40)(20,40)
\ArrowLine(0,0)(20,0)
\ArrowLine(80,40)(60,40)
\ArrowLine(80,0)(60,0)
\DashLine(20,40)(60,40)3
\DashLine(20,0)(60,0)3
\ArrowLine(20,20)(20,40)
\ArrowLine(20,20)(20,0)
\Line(17,17)(23,23)
\Line(17,23)(23,17)
\ArrowLine(60,20)(60,40)
\ArrowLine(60,20)(60,0)
\Line(57,17)(63,23)
\Line(57,23)(63,17)
\Text(10,30)[]{\trgor}
\Text(10,10)[]{\trdol}
\Text(40,30)[]{\sfgor}
\Text(40,10)[]{\sfdol}
\Text(75,30)[]{\spgor}
\Text(75,10)[]{\spdol}
\Text(-10,40)[]{\iena}
\Text(-10,0)[]{\idva}
\Text(90,0)[]{\itri}
\Text(90,40)[]{\isti}
\end{picture}

\propto\;\;\;\;\;
(U^T\underline A\tilde D)_{13}(\tilde D^\dagger Y_D^* D_c^*)_{31,32}
(U_c^\dagger Y_U^\dagger\tilde U^*)_{13}(\tilde U^T\underline CE)_{3i}
\end{equation}

\begin{equation}
\def\tr{$\bar T$}
\def\sfgor{$\tilde t$}
\def\sfdol{$\tilde b$}
\def\spgor{$\tilde h_0^\dagger$}
\def\spdol{$\tilde{\bar h}_0^\dagger$}
\def\iena{$u$}
\def\idva{$e_i$}
\def\itri{${\bar d}_{1,2}^c$}
\def\isti{${\bar u}^c$}
\begin{picture}(80,40)(40,20)
\SetWidth{0.8}
\ArrowLine(0,40)(10,20)
\ArrowLine(0,0)(10,20)
\ArrowLine(80,40)(60,40)
\ArrowLine(80,0)(60,0)
\DashLine(10,20)(40,20)3
\DashLine(40,20)(60,40)3
\DashLine(40,20)(60,0)3
\ArrowLine(60,20)(60,40)
\ArrowLine(60,20)(60,0)
\Line(57,17)(63,23)
\Line(57,23)(63,17)
\Text(25,30)[]{\tr}
\Text(45,35)[]{\sfgor}
\Text(45,5)[]{\sfdol}
\Text(75,30)[]{\spgor}
\Text(75,10)[]{\spdol}
\Text(-10,40)[]{\iena}
\Text(-10,0)[]{\idva}
\Text(90,0)[]{\itri}
\Text(90,40)[]{\isti}
\end{picture}

\propto\;\;\;\;\;
(U^T\underline CE)_{1i}(D_c^\dagger Y_D^\dagger\tilde D^*)_{13,23}
(\tilde D^T\underline A\tilde U)_{33}(\tilde U^\dagger Y_U^*U_c^*)_{31}
\end{equation}

\begin{equation}
\def\trgor{$\tilde{\bar T}$}
\def\trdol{$\tilde T$}
\def\sfgor{$\tilde t^c$}
\def\sfdol{$\tilde\tau^c$}
\def\spgor{$\tilde h_0$}
\def\spdol{$\tilde{\bar h}_0$}
\def\iena{${\bar d}_{1,2}^c$}
\def\idva{${\bar u}^c$}
\def\itri{${\bar e}_i$}
\def\isti{$u$}
\begin{picture}(80,40)(40,20)
\SetWidth{0.8}
\ArrowLine(0,40)(20,40)
\ArrowLine(0,0)(20,0)
\ArrowLine(80,40)(60,40)
\ArrowLine(80,0)(60,0)
\DashLine(20,40)(60,40)3
\DashLine(20,0)(60,0)3
\ArrowLine(20,20)(20,40)
\ArrowLine(20,20)(20,0)
\Line(17,17)(23,23)
\Line(17,23)(23,17)
\ArrowLine(60,20)(60,40)
\ArrowLine(60,20)(60,0)
\Line(57,17)(63,23)
\Line(57,23)(63,17)
\Text(10,30)[]{\trgor}
\Text(10,10)[]{\trdol}
\Text(40,30)[]{\sfgor}
\Text(40,10)[]{\sfdol}
\Text(75,30)[]{\spgor}
\Text(75,10)[]{\spdol}
\Text(-10,40)[]{\iena}
\Text(-10,0)[]{\idva}
\Text(90,0)[]{\itri}
\Text(90,40)[]{\isti}
\end{picture}

\propto\;\;\;\;\;
(D_c^\dagger\underline D^\dagger\tilde U_c^*)_{13,23}
(\tilde U_c^TY_U^TU)_{31}(E^TY_E^T\tilde E_c)_{i3}
(\tilde E_c^\dagger\underline B^\dagger U_c^*)_{31}
\end{equation}

\begin{equation}
\def\tr{$\bar T$}
\def\sfgor{$\tilde t^c$}
\def\sfdol{$\tilde\tau^c$}
\def\spgor{$\tilde h_0$}
\def\spdol{$\tilde{\bar h}_0$}
\def\iena{${\bar d}_{1,2}^c$}
\def\idva{${\bar u}^c$}
\def\itri{${\bar e}_i$}
\def\isti{$u$}
\begin{picture}(80,40)(40,20)
\SetWidth{0.8}
\ArrowLine(0,40)(10,20)
\ArrowLine(0,0)(10,20)
\ArrowLine(80,40)(60,40)
\ArrowLine(80,0)(60,0)
\DashLine(10,20)(40,20)3
\DashLine(40,20)(60,40)3
\DashLine(40,20)(60,0)3
\ArrowLine(60,20)(60,40)
\ArrowLine(60,20)(60,0)
\Line(57,17)(63,23)
\Line(57,23)(63,17)
\Text(25,30)[]{\tr}
\Text(45,35)[]{\sfgor}
\Text(45,5)[]{\sfdol}
\Text(75,30)[]{\spgor}
\Text(75,10)[]{\spdol}
\Text(-10,40)[]{\iena}
\Text(-10,0)[]{\idva}
\Text(90,0)[]{\itri}
\Text(90,40)[]{\isti}
\end{picture}

\propto\;\;\;\;\;
(D_c^\dagger\underline D^\dagger U_c^*)_{11,21}
(E^TY_E^T\tilde E_c)_{i3}(\tilde E_c^\dagger\underline B^\dagger
\tilde U_c^*)_{33}(\tilde U_c^TY_U^TU)_{31}
\end{equation}

\begin{equation}
\def\trgor{$\tilde{\bar T}$}
\def\trdol{$\tilde T$}
\def\sfgor{$\tilde b^c$}
\def\sfdol{$\tilde t^c$}
\def\spgor{$\tilde{\bar h}_0$}
\def\spdol{$\tilde h_0$}
\def\iena{${\bar u}^c$}
\def\idva{${\bar e}_i^c$}
\def\itri{$u$}
\def\isti{$d_{1,2}$}
\begin{picture}(80,40)(40,20)
\SetWidth{0.8}
\ArrowLine(0,40)(20,40)
\ArrowLine(0,0)(20,0)
\ArrowLine(80,40)(60,40)
\ArrowLine(80,0)(60,0)
\DashLine(20,40)(60,40)3
\DashLine(20,0)(60,0)3
\ArrowLine(20,20)(20,40)
\ArrowLine(20,20)(20,0)
\Line(17,17)(23,23)
\Line(17,23)(23,17)
\ArrowLine(60,20)(60,40)
\ArrowLine(60,20)(60,0)
\Line(57,17)(63,23)
\Line(57,23)(63,17)
\Text(10,30)[]{\trgor}
\Text(10,10)[]{\trdol}
\Text(40,30)[]{\sfgor}
\Text(40,10)[]{\sfdol}
\Text(75,30)[]{\spgor}
\Text(75,10)[]{\spdol}
\Text(-10,40)[]{\iena}
\Text(-10,0)[]{\idva}
\Text(90,0)[]{\itri}
\Text(90,40)[]{\isti}
\end{picture}

\propto\;\;\;\;\;
(U_c^\dagger\underline D^*\tilde D_c^*)_{13}
(\tilde D_c^TY_D^TD)_{31,32}(U^TY_U\tilde U_c)_{13}
(\tilde U_c^\dagger\underline B^* E_c^*)_{3i}
\end{equation}

\begin{equation}
\def\tr{$T$}
\def\sfgor{$\tilde t^c$}
\def\sfdol{$\tilde b^c$}
\def\spgor{$\tilde h_0$}
\def\spdol{$\tilde{\bar h}_0$}
\def\iena{${\bar e}_i^c$}
\def\idva{${\bar u}^c$}
\def\itri{$d_{1,2}$}
\def\isti{$u$}
\begin{picture}(80,40)(40,20)
\SetWidth{0.8}
\ArrowLine(0,40)(10,20)
\ArrowLine(0,0)(10,20)
\ArrowLine(80,40)(60,40)
\ArrowLine(80,0)(60,0)
\DashLine(10,20)(40,20)3
\DashLine(40,20)(60,40)3
\DashLine(40,20)(60,0)3
\ArrowLine(60,20)(60,40)
\ArrowLine(60,20)(60,0)
\Line(57,17)(63,23)
\Line(57,23)(63,17)
\Text(25,30)[]{\tr}
\Text(45,35)[]{\sfgor}
\Text(45,5)[]{\sfdol}
\Text(75,30)[]{\spgor}
\Text(75,10)[]{\spdol}
\Text(-10,40)[]{\iena}
\Text(-10,0)[]{\idva}
\Text(90,0)[]{\itri}
\Text(90,40)[]{\isti}
\end{picture}

\propto\;\;\;\;\;
(E_c^\dagger\underline B^\dagger U_c^*)_{i1}(D^TY_D\tilde D_c)_{13,23}
(\tilde D_c^\dagger\underline D^\dagger\tilde U_c^*)_{33}
(\tilde U_c^TY_U^TU)_{31}
\end{equation}

\begin{equation}
\def\trgor{$\tilde{\bar T}$}
\def\trdol{$\tilde T$}
\def\sfgor{$\tilde t$}
\def\sfdol{$\tilde\tau$}
\def\spgor{$\tilde h_0^\dagger$}
\def\spdol{$\tilde{\bar h}_0^\dagger$}
\def\iena{$d_{1,2}$}
\def\idva{$u$}
\def\itri{${\bar e}_i^c$}
\def\isti{${\bar u}^c$}
\begin{picture}(80,40)(40,20)
\SetWidth{0.8}
\ArrowLine(0,40)(20,40)
\ArrowLine(0,0)(20,0)
\ArrowLine(80,40)(60,40)
\ArrowLine(80,0)(60,0)
\DashLine(20,40)(60,40)3
\DashLine(20,0)(60,0)3
\ArrowLine(20,20)(20,40)
\ArrowLine(20,20)(20,0)
\Line(17,17)(23,23)
\Line(17,23)(23,17)
\ArrowLine(60,20)(60,40)
\ArrowLine(60,20)(60,0)
\Line(57,17)(63,23)
\Line(57,23)(63,17)
\Text(10,30)[]{\trgor}
\Text(10,10)[]{\trdol}
\Text(40,30)[]{\sfgor}
\Text(40,10)[]{\sfdol}
\Text(75,30)[]{\spgor}
\Text(75,10)[]{\spdol}
\Text(-10,40)[]{\iena}
\Text(-10,0)[]{\idva}
\Text(90,0)[]{\itri}
\Text(90,40)[]{\isti}
\end{picture}

\propto\;\;\;\;\;
(D^T\underline A\tilde U)_{13,23}
(\tilde U^\dagger Y_U^*U_c^*)_{31}(E_c^\dagger Y_E^*\tilde E^*)_{i3}
(\tilde E^T\underline C^T U)_{31}
\end{equation}

\begin{equation}
\def\tr{$T$}
\def\sfgor{$\tilde\tau$}
\def\sfdol{$\tilde t$}
\def\spgor{$\tilde{\bar h}_0^\dagger$}
\def\spdol{$\tilde h_0^\dagger$}
\def\iena{$u$}
\def\idva{$d_{1,2}$}
\def\itri{${\bar u}^c$}
\def\isti{${\bar e}_i^c$}
\begin{picture}(80,40)(40,20)
\SetWidth{0.8}
\ArrowLine(0,40)(10,20)
\ArrowLine(0,0)(10,20)
\ArrowLine(80,40)(60,40)
\ArrowLine(80,0)(60,0)
\DashLine(10,20)(40,20)3
\DashLine(40,20)(60,40)3
\DashLine(40,20)(60,0)3
\ArrowLine(60,20)(60,40)
\ArrowLine(60,20)(60,0)
\Line(57,17)(63,23)
\Line(57,23)(63,17)
\Text(25,30)[]{\tr}
\Text(45,35)[]{\sfgor}
\Text(45,5)[]{\sfdol}
\Text(75,30)[]{\spgor}
\Text(75,10)[]{\spdol}
\Text(-10,40)[]{\iena}
\Text(-10,0)[]{\idva}
\Text(90,0)[]{\itri}
\Text(90,40)[]{\isti}
\end{picture}

\propto\;\;\;\;\;
(U^T\underline AD)_{11,12}
(U_c^\dagger Y_U^\dagger\tilde U^*)_{13}(\tilde U^T\underline
C\tilde E)_{33}(\tilde E^\dagger Y_E^\dagger E_c^*)_{3i}
\end{equation}

\begin{equation}
\def\trgor{$\tilde T$}
\def\trdol{$\tilde{\bar T}$}
\def\sfgor{$\tilde b$}
\def\sfdol{$\tilde t$}
\def\spgor{$\tilde V_0$}
\def\spdol{$\tilde V_0$}
\def\iena{$u$}
\def\idva{$e_i$}
\def\itri{$u$}
\def\isti{$d_{1,2}$}
\begin{picture}(80,40)(40,20)
\SetWidth{0.8}
\ArrowLine(0,40)(20,40)
\ArrowLine(0,0)(20,0)
\ArrowLine(80,40)(60,40)
\ArrowLine(80,0)(60,0)
\DashLine(20,40)(60,40)3
\DashLine(20,0)(60,0)3
\ArrowLine(20,20)(20,40)
\ArrowLine(20,20)(20,0)
\Line(17,17)(23,23)
\Line(17,23)(23,17)
\ArrowLine(60,20)(60,40)
\ArrowLine(60,20)(60,0)
\Line(57,17)(63,23)
\Line(57,23)(63,17)
\Text(10,30)[]{\trgor}
\Text(10,10)[]{\trdol}
\Text(40,30)[]{\sfgor}
\Text(40,10)[]{\sfdol}
\Text(75,30)[]{\spgor}
\Text(75,10)[]{\spdol}
\Text(-10,40)[]{\iena}
\Text(-10,0)[]{\idva}
\Text(90,0)[]{\itri}
\Text(90,40)[]{\isti}
\end{picture}

\propto\;\;\;\;\;
(U^T\underline A\tilde D)_{13}(\tilde D^\dagger D)_{31,32}
(U^T\tilde U^*)_{13}(\tilde U^T\underline CE)_{3i}
\end{equation}

\begin{equation}
\def\tr{$\bar T$}
\def\sfgor{$\tilde b$}
\def\sfdol{$\tilde t$}
\def\spgor{$\tilde V_0$}
\def\spdol{$\tilde V_0$}
\def\iena{$u$}
\def\idva{$e_i$}
\def\itri{$u$}
\def\isti{$d_{1,2}$}
\begin{picture}(80,40)(40,20)
\SetWidth{0.8}
\ArrowLine(0,40)(10,20)
\ArrowLine(0,0)(10,20)
\ArrowLine(80,40)(60,40)
\ArrowLine(80,0)(60,0)
\DashLine(10,20)(40,20)3
\DashLine(40,20)(60,40)3
\DashLine(40,20)(60,0)3
\ArrowLine(60,20)(60,40)
\ArrowLine(60,20)(60,0)
\Line(57,17)(63,23)
\Line(57,23)(63,17)
\Text(25,30)[]{\tr}
\Text(45,35)[]{\sfgor}
\Text(45,5)[]{\sfdol}
\Text(75,30)[]{\spgor}
\Text(75,10)[]{\spdol}
\Text(-10,40)[]{\iena}
\Text(-10,0)[]{\idva}
\Text(90,0)[]{\itri}
\Text(90,40)[]{\isti}
\end{picture}

\propto\;\;\;\;\;
(U^T\underline CE)_{1i}(U^T\tilde U^*)_{13}
(\tilde U^T\underline A\tilde D)_{33}(\tilde D^\dagger D)_{31,32}
\end{equation}

\begin{equation}
\def\trgor{$\tilde{\bar T}$}
\def\trdol{$\tilde T$}
\def\sfgor{$\tilde\tau$}
\def\sfdol{$\tilde t$}
\def\spgor{$\tilde V_0$}
\def\spdol{$\tilde V_0$}
\def\iena{$u$}
\def\idva{$d_{1,2}$}
\def\itri{$u$}
\def\isti{$e_i$}
\begin{picture}(80,40)(40,20)
\SetWidth{0.8}
\ArrowLine(0,40)(20,40)
\ArrowLine(0,0)(20,0)
\ArrowLine(80,40)(60,40)
\ArrowLine(80,0)(60,0)
\DashLine(20,40)(60,40)3
\DashLine(20,0)(60,0)3
\ArrowLine(20,20)(20,40)
\ArrowLine(20,20)(20,0)
\Line(17,17)(23,23)
\Line(17,23)(23,17)
\ArrowLine(60,20)(60,40)
\ArrowLine(60,20)(60,0)
\Line(57,17)(63,23)
\Line(57,23)(63,17)
\Text(10,30)[]{\trgor}
\Text(10,10)[]{\trdol}
\Text(40,30)[]{\sfgor}
\Text(40,10)[]{\sfdol}
\Text(75,30)[]{\spgor}
\Text(75,10)[]{\spdol}
\Text(-10,40)[]{\iena}
\Text(-10,0)[]{\idva}
\Text(90,0)[]{\itri}
\Text(90,40)[]{\isti}
\end{picture}

\propto\;\;\;\;\;
(U^T\underline C\tilde E)_{13}(\tilde E^\dagger E)_{3i}
(U^T\tilde U^*)_{13}(\tilde U^T\underline AD)_{31,32}
\end{equation}

\begin{equation}
\def\tr{$T$}
\def\sfgor{$\tilde\tau$}
\def\sfdol{$\tilde t$}
\def\spgor{$\tilde V_0$}
\def\spdol{$\tilde V_0$}
\def\iena{$u$}
\def\idva{$d_{1,2}$}
\def\itri{$u$}
\def\isti{$e_i$}
\begin{picture}(80,40)(40,20)
\SetWidth{0.8}
\ArrowLine(0,40)(10,20)
\ArrowLine(0,0)(10,20)
\ArrowLine(80,40)(60,40)
\ArrowLine(80,0)(60,0)
\DashLine(10,20)(40,20)3
\DashLine(40,20)(60,40)3
\DashLine(40,20)(60,0)3
\ArrowLine(60,20)(60,40)
\ArrowLine(60,20)(60,0)
\Line(57,17)(63,23)
\Line(57,23)(63,17)
\Text(25,30)[]{\tr}
\Text(45,35)[]{\sfgor}
\Text(45,5)[]{\sfdol}
\Text(75,30)[]{\spgor}
\Text(75,10)[]{\spdol}
\Text(-10,40)[]{\iena}
\Text(-10,0)[]{\idva}
\Text(90,0)[]{\itri}
\Text(90,40)[]{\isti}
\end{picture}

\propto\;\;\;\;\;
(U^T\underline AD)_{11,12}(U^T\tilde U^*)_{13}
(\tilde U^T\underline C\tilde E)_{33}(\tilde E^\dagger E)_{3i}
\end{equation}

\begin{equation}
\def\trgor{$\tilde{\bar T}^\dagger$}
\def\trdol{$\tilde T^\dagger$}
\def\sfgor{$\tilde b^c$}
\def\sfdol{$\tilde t^c$}
\def\spgor{$\tilde V_0^\dagger$}
\def\spdol{$\tilde V_0^\dagger$}
\def\iena{${\bar u}^c$}
\def\idva{${\bar e}_i^c$}
\def\itri{${\bar u}^c$}
\def\isti{${\bar d}_{1,2}^c$}
\begin{picture}(80,40)(40,20)
\SetWidth{0.8}
\ArrowLine(0,40)(20,40)
\ArrowLine(0,0)(20,0)
\ArrowLine(80,40)(60,40)
\ArrowLine(80,0)(60,0)
\DashLine(20,40)(60,40)3
\DashLine(20,0)(60,0)3
\ArrowLine(20,20)(20,40)
\ArrowLine(20,20)(20,0)
\Line(17,17)(23,23)
\Line(17,23)(23,17)
\ArrowLine(60,20)(60,40)
\ArrowLine(60,20)(60,0)
\Line(57,17)(63,23)
\Line(57,23)(63,17)
\Text(10,30)[]{\trgor}
\Text(10,10)[]{\trdol}
\Text(40,30)[]{\sfgor}
\Text(40,10)[]{\sfdol}
\Text(75,30)[]{\spgor}
\Text(75,10)[]{\spdol}
\Text(-10,40)[]{\iena}
\Text(-10,0)[]{\idva}
\Text(90,0)[]{\itri}
\Text(90,40)[]{\isti}
\end{picture}

\propto\;\;\;\;\;
(U_c^\dagger\underline D^*\tilde D_c^*)_{13}(\tilde D_c^T D_c^*)_{31,32}
(U_c^\dagger\tilde U_c)_{13}(\tilde U_c^\dagger\underline B^*E_c^*)_{3i}
\end{equation}

\begin{equation}
\def\tr{$T$}
\def\sfgor{$\tilde b^c$}
\def\sfdol{$\tilde t^c$}
\def\spgor{$\tilde V_0^\dagger$}
\def\spdol{$\tilde V_0^\dagger$}
\def\iena{${\bar u}^c$}
\def\idva{${\bar e}_i^c$}
\def\itri{${\bar u}^c$}
\def\isti{${\bar d}_{1,2}^c$}
\begin{picture}(80,40)(40,20)
\SetWidth{0.8}
\ArrowLine(0,40)(10,20)
\ArrowLine(0,0)(10,20)
\ArrowLine(80,40)(60,40)
\ArrowLine(80,0)(60,0)
\DashLine(10,20)(40,20)3
\DashLine(40,20)(60,40)3
\DashLine(40,20)(60,0)3
\ArrowLine(60,20)(60,40)
\ArrowLine(60,20)(60,0)
\Line(57,17)(63,23)
\Line(57,23)(63,17)
\Text(25,30)[]{\tr}
\Text(45,35)[]{\sfgor}
\Text(45,5)[]{\sfdol}
\Text(75,30)[]{\spgor}
\Text(75,10)[]{\spdol}
\Text(-10,40)[]{\iena}
\Text(-10,0)[]{\idva}
\Text(90,0)[]{\itri}
\Text(90,40)[]{\isti}
\end{picture}

\propto\;\;\;\;\;
(U_c^\dagger\underline B^*E_c^*)_{1i}(U_c^\dagger\tilde U_c)_{13}
(\tilde U_c^\dagger\underline D^*\tilde D_c^*)_{33}
(\tilde D_c^TD_c^*)_{31,32}
\end{equation}

\begin{equation}
\def\trgor{$\tilde T^\dagger$}
\def\trdol{$\tilde{\bar T}^\dagger$}
\def\sfgor{$\tilde\tau^c$}
\def\sfdol{$\tilde t^c$}
\def\spgor{$\tilde V_0^\dagger$}
\def\spdol{$\tilde V_0^\dagger$}
\def\iena{${\bar u}^c$}
\def\idva{${\bar d}_{1,2}^c$}
\def\itri{${\bar u}^c$}
\def\isti{${\bar e}_i^c$}
\begin{picture}(80,40)(40,20)
\SetWidth{0.8}
\ArrowLine(0,40)(20,40)
\ArrowLine(0,0)(20,0)
\ArrowLine(80,40)(60,40)
\ArrowLine(80,0)(60,0)
\DashLine(20,40)(60,40)3
\DashLine(20,0)(60,0)3
\ArrowLine(20,20)(20,40)
\ArrowLine(20,20)(20,0)
\Line(17,17)(23,23)
\Line(17,23)(23,17)
\ArrowLine(60,20)(60,40)
\ArrowLine(60,20)(60,0)
\Line(57,17)(63,23)
\Line(57,23)(63,17)
\Text(10,30)[]{\trgor}
\Text(10,10)[]{\trdol}
\Text(40,30)[]{\sfgor}
\Text(40,10)[]{\sfdol}
\Text(75,30)[]{\spgor}
\Text(75,10)[]{\spdol}
\Text(-10,40)[]{\iena}
\Text(-10,0)[]{\idva}
\Text(90,0)[]{\itri}
\Text(90,40)[]{\isti}
\end{picture}

\propto\;\;\;\;\;
(U_c^\dagger\underline B^*\tilde E_c^*)_{13}(\tilde E_c^TE_c^*)_{3i}
(U_c^\dagger\tilde U_c)_{13}
(\tilde U_c^\dagger\underline D^*D_c^*)_{31,32}
\end{equation}

\begin{equation}
\def\tr{$\bar T$}
\def\sfgor{$\tilde\tau^c$}
\def\sfdol{$\tilde t^c$}
\def\spgor{$\tilde V_0^\dagger$}
\def\spdol{$\tilde V_0^\dagger$}
\def\iena{${\bar u}^c$}
\def\idva{${\bar d}_{1,2}^c$}
\def\itri{${\bar u}^c$}
\def\isti{${\bar e}_i^c$}
\begin{picture}(80,40)(40,20)
\SetWidth{0.8}
\ArrowLine(0,40)(10,20)
\ArrowLine(0,0)(10,20)
\ArrowLine(80,40)(60,40)
\ArrowLine(80,0)(60,0)
\DashLine(10,20)(40,20)3
\DashLine(40,20)(60,40)3
\DashLine(40,20)(60,0)3
\ArrowLine(60,20)(60,40)
\ArrowLine(60,20)(60,0)
\Line(57,17)(63,23)
\Line(57,23)(63,17)
\Text(25,30)[]{\tr}
\Text(45,35)[]{\sfgor}
\Text(45,5)[]{\sfdol}
\Text(75,30)[]{\spgor}
\Text(75,10)[]{\spdol}
\Text(-10,40)[]{\iena}
\Text(-10,0)[]{\idva}
\Text(90,0)[]{\itri}
\Text(90,40)[]{\isti}
\end{picture}

\propto\;\;\;\;\;
(U_c^\dagger\underline D^*\tilde D_c^*)_{11,12}
(U_c^\dagger\tilde U_c)_{13}
(\tilde U_c^\dagger\underline B^*\tilde E_c^*)_{33}
(\tilde E_c^T E_c^*)_{3i}
\end{equation}

\end{document}